\documentclass[preprint,12pt]{elsarticle}

\usepackage{amssymb}
\journal{Physica A}

\begin{document}

\begin{frontmatter}



\title{The Effect of Modern Traffic Information on Braess' Paradox}


\author{Stefan Bittihn, Andreas Schadschneider\footnote{Corresponding author;
email address: as@thp.uni-koeln.de}}

\address{Institute for Theoretical Physics, Universit\"at zu K\"oln,
  50937 K\"oln, Germany}

\begin{abstract}
Braess' paradox has been shown to appear rather generically in many
systems of transport on networks. It is especially relevant for
vehicular traffic where it shows that in certain situations building
a new road in an urban or highway network can lead to increased
average travel times for {\it all} users. Here we address the
question whether this changes if the drivers (agents) have access to
traffic information as available for modern traffic networks, i.e.
through navigation apps and or personal experiences in the past. We
study the effect of traffic information in the classical Braess
network, but using a microscopic model for the traffic dynamics, to
find out if the paradox can really be observed in such a scenario or
if it only exists in some theoretically available user optima that
are never realized by drivers that base their route choice decisions
intelligently upon realistic traffic information. We address this
question for different splits of the two information types.
\end{abstract}




\end{frontmatter}


\section{Introduction}
\label{sec-intro}

Since its discovery in D. Braess' proof-of-concept paper in
1968~\cite{braess68,braessnw05}, Braess' paradox has become a
well-known phenomenon, first in traffic science and later in other
fields as well, e.g. in mechanical and electrical
systems~\cite{cohen1991paradoxical,penchina2003}, oscillator
networks, power grids and biological
networks~\cite{witthaut2012braess,tchuisseu2018,motter2018antagonistic},
microfluidic networks~\cite{case2019}, and pedestrian
dynamics~\cite{crociani2016}. Braess' paradox states that in
(congested) road networks used by selfish drivers under certain
conditions the addition of a new road can lead to higher travel
times for all network users. The paradox is established on the
assumption that networks of selfish users evolve into stable states,
the so-called user optimum states~\cite{wardrop1952}. This insight
has later been generalized to the concept of "price of
anarchy"~\cite{youn2008}. The paradox has been generalized in
various
aspects~\cite{stewart1980equilibrium,pas1997braess,murchland1970braess,steinberg83,frank1981braess,dafermos1984some,nagurney2010negation}.
Most of this research is, as the original work, built upon
simplified macroscopic mathematical models of traffic flow. It is
worth mentioning that the paradox has been observed in real world
traffic networks~\cite{baker2009,kolata1990,vidal2006}.

In the spirit of Dietrich Stauffer we consider Braess' paradox as a
problem of "exotic statistical physics" \cite{stauffer}. In two
recent papers~\cite{bittihn2016,bittihn2018} we studied the paradox
in traffic networks with a more realistic model of traffic flow,
i.e. the totally asymmetric exclusion process
(TASEP)~\cite{schadschneider2010stochastic,Schuetz-review,BlytheE07}.
For these cases we could show that it can also be observed and
indeed occupies large parts of the phase space.

In most of the previous studies the paradox is observed in the
following way: it is shown that in the traffic networks with and
without an added road user optima exist for the same total demand
(total numbers of cars using the network) and that in some cases the
user optima of the networks with the new road have higher travel
times than those of the networks without the new road. Furthermore,
in our previous work \cite{bittihn2016,bittihn2018} on the paradox
in TASEP networks we dealt with systems that evolved into a
stationary state and considered stationary travel times only,
neglecting the evolution into these states. In our recent work
\cite{bittihn2020} we have started to look at the effects of traffic
information on the states realized in Braess-type road networks.

In the present paper we continue to consider a more microscopic
point-of-view and address the question if the attainable user optima
are actually realized if the network users (from here on also called
{\it agents}) make informed route choices based upon realistic
traffic information. The question whether user optima (and thus also
Braess' paradox) are realized in real world networks has been a long
standing discussion, examined in various scientific fields.
Foremost, studies from traffic scientists suggest, that travel time
minimization is not the only factor driving route
choices~\cite{parthasarathi2013,chen2001using,zhu2015}. In the field
of behavioral economics several experiments with real human
participants were
performed~\cite{rapoport2009choice,selten2007,meneguzzer2013day,ye2017}.
In these studies, generally the traffic flow is modelled by a
macroscopic mathematical approach. In the physics community, route
choice processes and the question whether user optima are realized
were mainly studied by so-called multi-agent
techniques~\cite{wahle2000decision,bazzan1999agents,bazzan2005case,he2014,levy2016emergence}:
those are mainly proof-of-concept simulations in which the traffic
flow is modelled by a stochastic microscopic model and the route
choices are performed equally by all agents based on certain types
of information.

In the present article we model the traffic flow by a microscopic
model and implement a route choice algorithm by which all individual
agents make 'intelligent' decisions. Users have access to
personal-historical information, i.e. their own memory of travel
times experienced in the past, or public-predictive information,
i.e. estimates of travel times based on the positions of all agents
at a given time. The latter approximates information from personal
navigational systems or smartphone apps, like the crowdsourced app
Google Maps~\cite{google-blog-maps}. In the present paper we study
the case in which both information types are present in the road
network. In \cite{bittihn2020} we considered situations in which
only one of the two types was available to each agent. We examined
four different points of the phase space. We could show that
personal-historical information realizes user optima in all four
states, while public-predictive information realizes stable user
optimum states at low densities and oscillations around attainable
user optima are observed at higher densities. Here we consider the
more realistic scenario in which different splits of agents have
access to the two kinds of information. Such a situation is relevant
for most real modern day traffic networks. Analysis of real world
GPS-traffic data hints at routing apps making travel time
minimization the most important route choice factor and also suggest
that these apps lead to realizations of user
optima~\cite{meneguzzer2013day,cabannes2018impact}, which has also
been backed by large-scale simulations~\cite{cabannes2018impact}.

%%%%%%%%%%%%%%%%%%%%%%%%%%%%%%%%%%%%%%%%%%%%%%%%%%%%%%%%%%%%%%%%%%%%%%%%%%%%%%

\section{The paradox}

A fundamental assumption underlying Braess' paradox is that the
users of the (traffic) network are selfish and that their egoistic
behavior drives the system into stable network states (user optima,
{\it uo}). In this context, a {\it network state} refers to the set
of strategies of all network users. A user's {\it strategy} denotes
how the user chooses their route. A {\it route} refers to a
connection between an origin and a destination and can be comprised
of various {\it roads} or, in network terminology, edges that can be
connected through {\it junction sites}. In the following we only
consider situations in which all users want to go from the same
origin to the same destination. The {\it user optimum} is realized
if the user's route choices (and thus their distribution onto the
roads of the network) lead to equal travel times on all used routes,
which are lower than travel times on any unused
routes~\cite{wardrop1952}. For a {\it pure strategy} a user chooses
exactly one route whereas for a {\it mixed strategy} she chooses a
route according to probabilities assigned to all routes.
Accordingly, if all agents use pure strategies, {\it pure user
optima} ({\it puo}) are observed, while in the case of all users
using mixed strategies  {\it mixed user optima} ({\it muo}) are
observed. In the latter case, the user optimum is realized if the
expectation values of the travel times of all used routes are equal
and lower than those of any unused routes. Since we are using a
traffic model with microscopic stochastic dynamics we have to
consider mean values of travel times both for pure and mixed user
optima.

\subsection{Braess' original example}

\begin{figure}[ht]
  \centering
  \includegraphics[width=0.6\columnwidth]{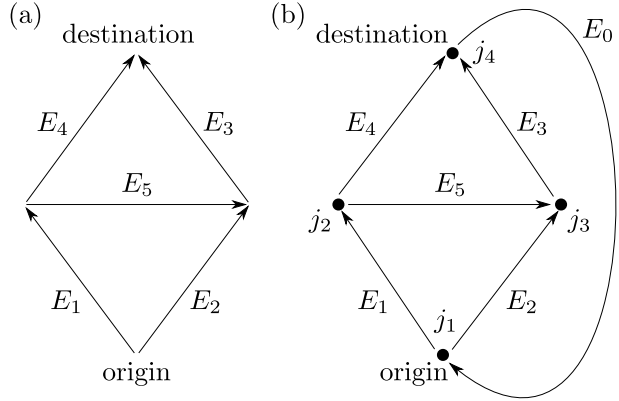}
  \caption{\label{fig:braess_network_of_taseps}
  The Braess network. Edge $E_5$   is considered to be the road that is added to the network.
  The network without/with
  $E_5$ is called the 4link/5link network. All cars are imagined to go from the same
  origin at the bottom to the same destination at the top. In the 4link network, cars
  can choose between the two routes 14 and 23. In the 5link, additionally the new
  route 153 exists, due to the addition of road $E_5$.
  (a) The network used in Braess' original work.
  (b) The network used in the present work:
  the edges $E_i$ consist of TASEPs of lengths $L_i$. We added the junction sites $j_k$
  and periodic boundary conditions through $E_0$.}
\end{figure}

Braess demonstrated the paradox~\cite{braess68,braessnw05} for the
simple network shown in Fig.~\ref{fig:braess_network_of_taseps}~(a).
It is comprised of the 5 network edges ${E_1,\dots,E_5}$. It is
assumed that all network users want to go from the origin at the
bottom to the destination at the top. Edge $E_5$ is considered to be
the new road that is added to the network. The network without/with
edge $E_5$ is from here on called the 4link/5link network and
corresponding variables are indicated by superscripts (4) and (5),
respectively. Routes in the network are labelled according to the
numbering scheme of the constituent edges. In the 4link network
there are two available routes from origin to destination: route 14
and route 23. The addition of edge $E_5$ realizes a new route, route
153.

In Braess' original work \cite{braess68,braessnw05} a macroscopic
traffic model was used with travel times determined by simple travel
time functions that increase linearly with the number of cars using
the road. The approach also neglected potential correlations that
exist between the different roads. Braess showed that for specific
choices of the number of agents and parameters in the travel time
functions an apparently paradox situation can occur where the travel
times in the 4link network's user optimum are smaller than those of
the 5link network's user optimum, i.e. adding an additional route or
edge to the network leads to an increase of travel times for all
(selfish) agents. In Braess' example both for the 4link and the
5link network unique pure optima exist. It was also shown that the
same effect can be observed with mixed user
optima~\cite{bittihn2020}. Several generalisations and additional
results about the paradox were obtained in the context of
macroscopic mathematical traffic
models~\cite{stewart1980equilibrium,pas1997braess,murchland1970braess,steinberg83,frank1981braess,dafermos1984some,nagurney2010negation}.

\subsection{The Braess network of TASEPs}

In \cite{bittihn2016,bittihn2018} we have considered the occurrence
of Braess' paradox for the same network structure, but with a more
realistic traffic model described by stochastic, microscopic
dynamics, namely the totally asymmetric exclusion process (TASEP).
In this situation user optima realizing the paradox can also be
observed.

In the model based on TASEP traffic dynamics each car or agent is
represented as a particle subject to hard core repulsion. A road is
modelled by discrete cells that can be either empty or occupied by
one particle. Using a random-sequential update, particles can only
move forward to an empty cell in front. Each edge (road) $E_i$ is
represented by a TASEP of length $L_i$ cells. As shown in
Fig.~\ref{fig:braess_network_of_taseps}~(b), we examined the same
network structure as Braess, with four junction sites
$j_1,\dots,j_4$ and edge $E_0$ ensuring periodic boundary conditions
and thus a constant total number of particles $M$. We could
show~\cite{bittihn2016,bittihn2018} that user optima leading to
Braess' paradox also exist in those TASEP networks. We demonstrated
this by {\it externally setting all agents' strategies}: the route
choices of the agents were controlled by global parameters. By
varying these parameters, we found attainable user optima of the
4link and 5link networks and compared their travel times (for the
same total numbers of particles $M$). This allowed us to determine
the phase diagrams that are presented in the following. Note, that
no individual, realistic route choice decisions were modelled. The
obtained phase diagrams are valid only if all agents follow the
externally set route choices.

\begin{figure}[ht]
  \centering
  \includegraphics[width=0.49\columnwidth]{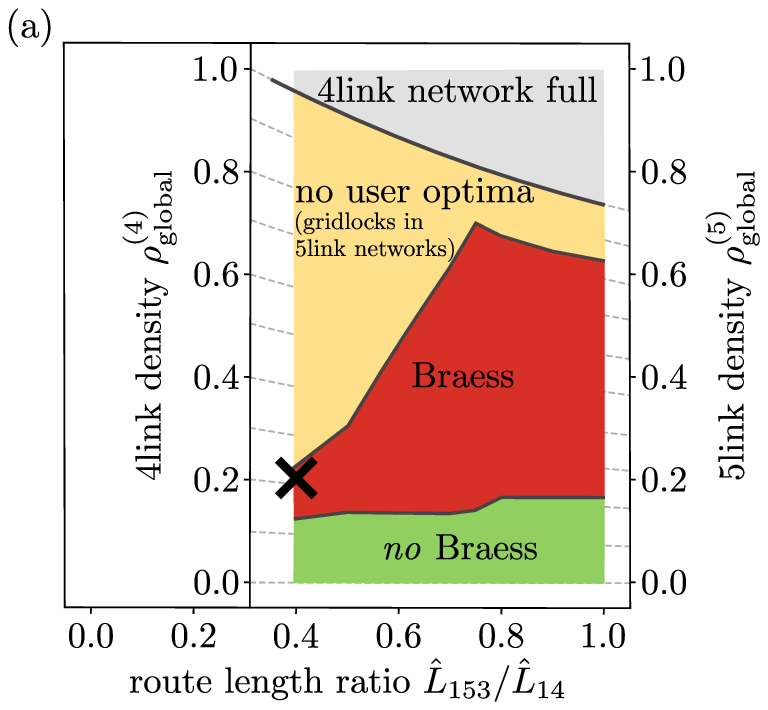}
  \includegraphics[width=0.49\columnwidth]{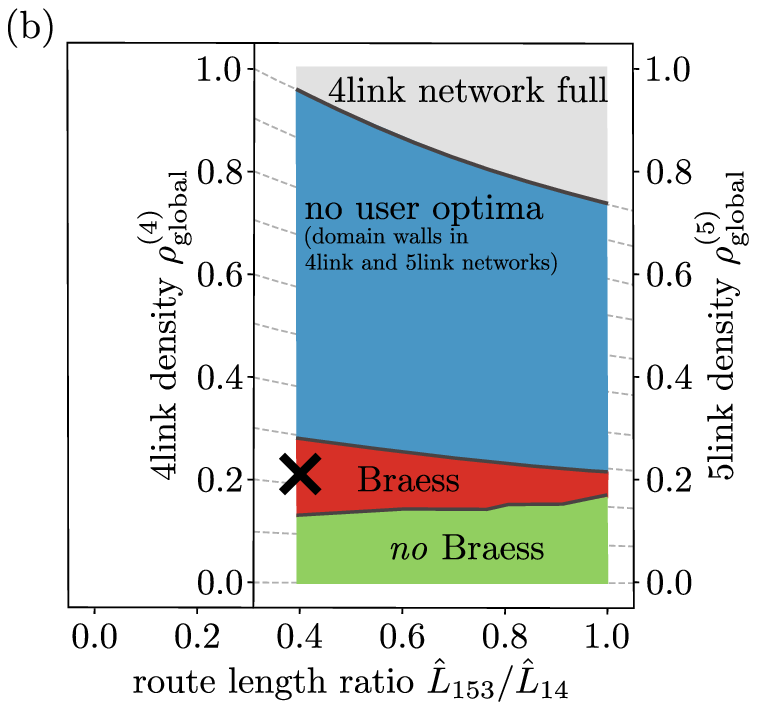}
  \caption{\label{fig:braess_phases_ext_tuned}Phase diagrams for Braess' network of TASEPs
  with externally tuned strategies and (a) fixed personal strategies and (b) turning probabilities.
  These phase diagrams are simplified. For full details (sub-phases etc.) the reader is referred
  to~\cite{bittihn2018,bittihn2016}.
  The route choice algorithm was tested on the state marked with a black $\mathrm{\times}$.}
\end{figure}

In~\cite{bittihn2018} we considered the specific case where all
agents follow fixed personal (i.e. pure) strategies. In this case
fixed numbers of $N_{14},~N_{23},~N_{153}$ agents follow routes 14,
23 and 153, respectively. The distribution of agents onto the routes
can also be characterized by the two variables
$n_{\mathrm{l}}^{(j_1)}=1-\frac{N_{23}}{M}$ and
$n_{\mathrm{l}}^{(j_2)}=\frac{N_{14}}{N_{14}+N_{153}}$. In the 4link
network, as $E_5$ does not exist, $N_{153}=0$. By varying these
numbers a rich phase diagram was found that is shown in
Fig.~\ref{fig:braess_phases_ext_tuned}~(a) in a simplified form.

In~\cite{bittihn2016} we considered agents using {\it mixed}
strategies on the same network. The agents decide for a route based
on turning probabilities on junction sites $j_1$ and $j_2$.
%In this case all particles choose their routes according to the same turning probabilities.
%\remS{An agent} \comS{Each agent o. All agents o. Agents}
%\anmS{(dann wird nochmal expliziter kalr, dass alle Teilchen hier gleich sind)}
Agents on $j_1$ turn to the left (onto $E_1$) with probability
$\gamma$ and to the right (onto $E_2$) with probability $1-\gamma$.
In the 5link system
%\remS{an agent} \comS{each agent o. all agents o. agents} sitting
agents on $j_2$ turn to the left (onto $E_4$) with probability
$\delta$ and to the right (onto $E_5$) with probability $1-\delta$.
In the 4link system $E_5$ does not exist and thus $\delta=1$. The
resulting phase diagram is shown in
Fig.~\ref{fig:braess_phases_ext_tuned}~(b) in a simplified form.

For the phase diagrams shown in
Fig.~\ref{fig:braess_phases_ext_tuned}, the lengths of the TASEPs
were chosen as $L_0=1$, $L_1=L_3=100$, $L_2=L_4=500$. The length of
the new road $L_5$ and the total number of agents in the system $M$
are varied. The state of the network depends on $L_5$ and $M$ which
in the phase diagrams are characterized by the route length ratio
$\hat{L}_{153}/\hat{L}_{14}$ (note that $\hat{L}_{14}=\hat{L}_{23}$)
and the global densities
$\rho^{(4)}_{\mathrm{global}}=M/(4+\sum_{i=0}^4 L_i)$ and
$\rho^{(5)}_{\mathrm{global}} =M/(4+\sum_{i=0}^5 L_i)$ in the 4link
and 5link networks.

The phase diagrams shown in Fig.~\ref{fig:braess_phases_ext_tuned}
are simplified compared to those presented in
\cite{bittihn2018,bittihn2016} to emphasize the features relevant
for the present study. It is just shown where the Braess paradox
occurs ("Braess" phase, i.e. user optima in the 5link have higher
travel times than those of the 4link network) and where it does not
occur ("\textit{no} Braess" phase, i.e. user optima in the 5link
have lower travel times than those of the 4link network). These
phases can be subdivided. For more details including the sub-phases
and explanations about the regions in which no user optima could be
found ("no user optima") the reader is referred
to~\cite{bittihn2018,bittihn2016}. It can be seen that the paradox
can be observed in large parts of the phase space.

In the following we try to answer the following question: are the
user optima attainable by externally tuning all agents' strategies
actually realized if the agents make their route choice decisions
intelligently, similar to real drivers in modern traffic networks.
In other words: Would the paradox really be observed or does it only
exists in some theoretically attainable user optima that would never
be realized by realistic drivers.

%%%%%%%%%%%%%%%%%%%%%%%%%%%%%%%%%%%%%%%%%%%%%%%%%%%%%%%%%%%%%%%%%%%%%%%%%%%%%%%%%%

\section{The route choice algorithm}

To answer the question whether the paradox is realized by
realistically deciding drivers we implemented a route choice
algorithm. Agents choose their routes according to two different
types of information: personal-historical and public-predictive
information. Personal-historical information is typically relevant
in a commuter scenario and basically represents the driver's memory.
The agents go from origin to destination several times and remember
how long it took them in the past on the available routes. Their
future route choice decisions are then made upon this basis.
Public-predictive information, on the other hand, is typically
provided by navigational systems or smartphone routing apps. This
type of information is based upon the {\it current} status of the
network (how many cars are on which roads right now). Often,
expected travel times for all available routes are presented to the
driver who then decides for the one with the shortest predicted
travel time.

We have implemented a route choice algorithm based upon which all
agents in the system make individual 'intelligent' route choice
decisions before beginning each new round (i.e. before jumping from
$E_0$ to $j_1$) and in some cases also during rounds (a {\it round}
refers to going once from origin ($j_1$) to destination ($j_4$)).
The "goal" of the agents is to generally choose the route with the
lowest travel time. Based on the traffic information available to
him/her, each agent estimates the expected travel times on all
available routes. These estimations are stored in the variables
$T_{i,\mathrm{info}},\,i\in\{14,23,153\}$. Generally agents then
choose the route on which the lowest travel time is expected. If
this choice requires a route switch compared to the previous round,
it is only performed if the expected time-save is significant: the
value of
$|T_{14,\mathrm{info}}-T_{23,\mathrm{info}}|+|T_{14,\mathrm{info}}
-T_{153,\mathrm{info}}|+|T_{23,\mathrm{info}}-T_{153,\mathrm{info}}|$
is calculated and only if its value is larger than $\Delta T$, the
switch occurs. The agents thus act {\it boundedly
rational}~\cite{mahmassani1987boundedly}. A stochastic component is
added to account for random events that could disturb those
boundedly rational decisions: with probability $1-p_{\mathrm{info}}$
a random route is chosen.

During individual rounds, changes can occur in the following cases.
If e.g. an agent in the 4link network that chose to take route 23
before the round is sitting on $j_1$ and cannot jump to the target
site (first site of $E_2$) since this site is occupied, (s)he may
re-decide routes. If $T_{23,\mathrm{info}}\geq T_{14,\mathrm{info}}$
(agent chose route 23 based on a random decision before the round),
(s)he will then immediately switch to route 14. If
$T_{23,\mathrm{info}}<T_{14,\mathrm{info}}$, the particle will keep
trying to jump onto $E_2$ for $\kappa_{j_1,\mathrm{thres}}$ times
the to-be-expected saved time on route 23, i.e. for
$\kappa_{j_1,\mathrm{thres}}\cdot(T_{14,\mathrm{info}}-T_{23,\mathrm{info}})$
time steps. If the waiting time exceeds this value, (s)he will
switch to route 14. The same is true for the inverse route choice
situation (i.e. switching from route 14 to 23). In the 5link
network, additionally an equivalent mechanism is in place that also
includes route switches if an agent is waiting on $j_2$, introducing
the variable $\kappa_{j_2,\mathrm{thres}}$.

Public-predictive information is introduced in the following way.
Each time an agent relying on this kind of information makes a route
choice decision, for each edge $E_i$ the sum of numbers of particles
occupying this edge $n_i$ is collected. This mirrors the
crowdsourcing of information as in apps like Google
Maps~\cite{google-blog-maps}. For each edge $i$ then an approximated
travel time is calculated as $T_{i,\mathrm{est.}}\approx
L_i/(1-\rho_i)$ with $\rho_i=n_i/L_i$. This approximative travel
time is the exact travel time of a TASEP segment with periodic
boundary conditions~\cite{bittihn2016}. From this the
$T_{i,\mathrm{info}}$ are calculated:
$T_{14,\mathrm{info}}=T_{1,\mathrm{est.}}+T_{4,\mathrm{est.}}$ and
$T_{23,\mathrm{info}}$ and $T_{153,\mathrm{info}}$ accordingly.

Agents using personal-historical information remember the travel
times of the routes they used in the last $c_{\mathrm{mem}}$ rounds.
To get their estimated travel time values $T_{i,\mathrm{info}}$ that
they base their decisions upon, they calculate the mean values of
the travel times experienced in the last $c_{\mathrm{mem}}$ rounds
for each route. Additionally, these users remember the last travel
time they experienced on each route, even if they did not use a
specific route in the last $c_{\mathrm{mem}}$ rounds.

For the results presented in the present article, the parameters of
the route choice algorithm were chosen as
\begin{eqnarray}
% p_{\mathrm{info}}&=0.9, \\
% \Delta T_{\mathrm{thres}}&=10, \\
% \kappa_{j_1,\mathrm{thres}}=\kappa_{j_2,\mathrm{thres}}&=0.1, \\
% c_{\mathrm{mem}}&=30\,.
  p_{\mathrm{info}}=0.9, \quad
 \Delta T_{\mathrm{thres}}=10, \quad
 \kappa_{j_1,\mathrm{thres}}=\kappa_{j_2,\mathrm{thres}}=0.1, \quad
 c_{\mathrm{mem}}=30\,.
\end{eqnarray}
For more details on the algorithm including pseudo-code
representation we refer to~\cite{bittihn2020,bittihn2018phd}.

In~\cite{bittihn2020} we looked at four exemplary points of the
phase diagrams (two points where a Braess phase is expected, two
points where no Braess is expected) and tested if the expected user
optima are realized by agents basing their route choices upon our
algorithm with {\it all agents having access to either
personal-historical or public-predictive information}. We could show
that if all agents decide based on personal-historical information
%\remS{the expected}
user optima that are attainable by externally tuning the agents
decisions are realized. If all agents decide based on
public-predictive information, at lower densities attainable user
optima are realized and at higher global densities they are realized
on average. The reliance on this kind of information leads to
oscillations around the expected optima. In situations with higher
global densities, we could show that agents relying on
public-predictive information actually lead the system into a Braess
state in a case were a lower travel time in the 5link system was
expected.

%%%%%%%%%%%%%%%%%%%%%%%%%%%%%%%%%%%%%%%%%%%%%%%%%%%%%%%%%%%%%%%%%%%%%%%%%%%%%%%%%%

\section{Results}

Here we consider a scenario in which different splits of the two
information types are used: certain parts of the agents have access
only to personal-historical information while the other agents only
have access to public-predictive information. This represents the
realistic scenario of a traffic network that is used partly by
commuters who "know" travel times from their personal experiences in
the past and partly by users that rely on their navigational apps
for route decisions.

We examine a test state marked by the black $\mathbf{\times}$ in the
phase diagrams in Fig.~\ref{fig:braess_phases_ext_tuned}. It has the
parameters $L_5=37$ and $M=248$ which correspond to
$\hat{L}_{153}/\hat{L}_{14}=0.4$,
$\rho_{\mathrm{global}}^{(4)}\approx 0.21$,
$\rho_{\mathrm{global}}^{(5)}= 0.2$.  For externally tuned route
choices this is a ``Braess" state both for the previously studied
cases of fixed personal strategies and fixed turning probabilities.
The user optima that can be found by externally tuning the
strategies are given in in Table~\ref{tab:optima}.
In~\cite{bittihn2016,bittihn2018} we found so-called {\it boundedly
rational user optima}. This means that the mean travel times on the
used routes are not exactly equal. This can not be completely
avoided since we deal with stochastic transport models. Since the
travel times are not exactly equal, we present the maximum travel
time $T_{\mathrm{max}}$ measured on all used routes in the user
optima. From here on all travel times are measured in numbers of
Monte Carlo (MC) sweeps.

\begin{table}[h]
\begin{center}
    \begin{tabular}{| c | c | c |}
     \hline
    user optimum &  strategy & max. travel time \\ \hline \hline
    4link, pure uo & $n_{\mathrm{l,\,puo}}^{(j_1)(4)}=0.5$ &  $T^{(4)}_{\mathrm{max,puo}}\approx 764$ \\ \hline
    4link, mixed uo & $\gamma^{(4)}_{\mathrm{muo}}=0.5$ & $T_{\mathrm{max,muo}}^{(4)}\approx 763$\\ \hline \hline
    5link, first pure uo & $n_{\mathrm{l,\,puo(i)}}^{(j_1)(5)}=0.5$ & $T_{\mathrm{max,puo(i)}}^{(5)}\approx 978$ \\
    & $n_{\mathrm{l,\,puo(i)}}^{(j_2)(5)}=0.0$ &\\ \hline
    5link, second pure uo & $n_{\mathrm{l,\,puo(ii)}}^{(j_1)(5)}=1.0$ & $T_{\mathrm{max,puo(ii)}}^{(5)}\approx 978$ \\
    & $n_{\mathrm{l,\,puo(ii)}}^{(j_2)(5)}=0.5$ &\\ \hline
    5link, mixed uo & $\gamma^{(5)}_{\mathrm{muo}}=0.87$ & $T_{\mathrm{max,muo(i)}}^{(5)}\approx 895$ \\
    & $\delta^{(5)}_{\mathrm{muo}}=0.1$ &\\ \hline
    \end{tabular}
 \caption{\label{tab:optima}The user optima found for our test state,
 as marked by the black $\times$ in Fig.~\ref{fig:braess_phases_ext_tuned}.
 The (externally tuned) strategies realizing the optima and the corresponding
 maximum travel times are given. In the 4link system, one pure and one mixed
 user optimum exist, while in the 5link two pure and one mixed user optimum exist.}
\end{center}
\end{table}
In the following the values provided in Table~\ref{tab:optima} are
used as references to the systems' behaviors when used by agents
making decisions based on the route choice algorithm.

\begin{figure*}
  \centering
  \includegraphics[width=\columnwidth]{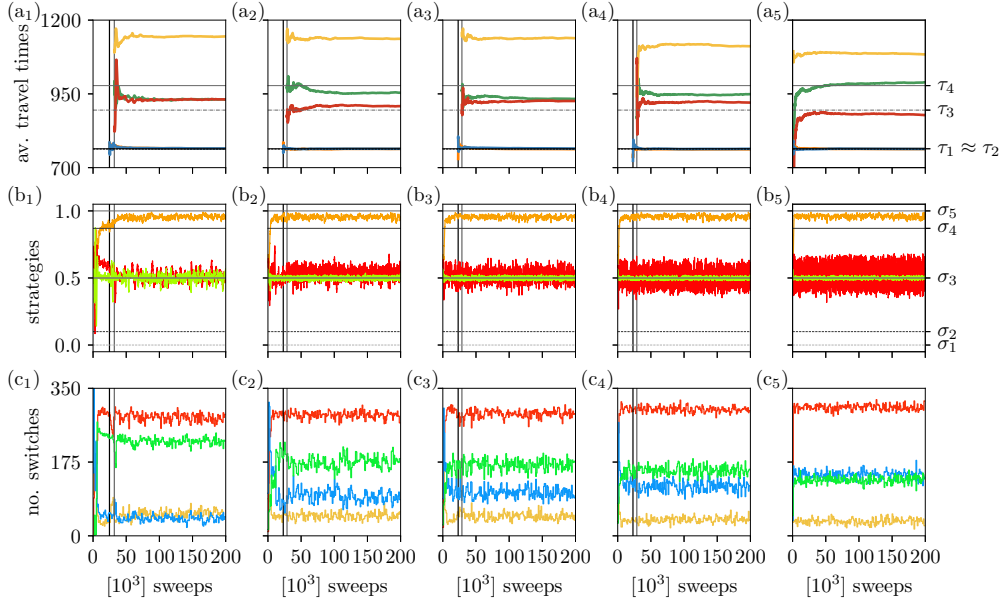}
  \caption{\label{fig:results}The results of the route choice algorithm. All plots share the same
  $x$-axes: the number of system sweeps (Monte Carlo sweeps) that were performed, i.e. the system
  time. All plots in one row share the same $y$-axes. The columns represent systems with varying
  amounts of agents using the two kinds of information. \textbf{Panels with indexes 1 to 5},
  i.e. columns from left to right, represent the following amounts of agents using personal-historical
  and public-predictive information: 100\% to 0\%, 75\% to 25\%, 50\% to 50\%, 25\% to 75\%, 0\% to 100\%.
  In systems with non-zero use of personal historical information the two vertical lines represent
  the system times the 4link system was relaxed (black vertical line) and the 5link system was
  relaxed (grey vertical line).
  \textbf{Panels (a$_i$)} show the average travel times measured on the routes. The orange and
  blue lines are the average travel times of routes 14 and 23 in the 4link system. The green, yellow,
  red lines are the average travel times of routes 14, 23, 153 in the 5link system. For comparison
  the thin grey horizontal lines show the travel times of the user optima obtained by externally
  tuning the strategies. Their values are indicated by the $\tau_i$ on the second $y$-axis of
  panel (a$_5$) with $T_{\rm max,muo}^{(4)}=\tau_1$, $T_{\rm max,puo}^{(4)}=\tau_2$,
  $T_{\rm max,muo}^{(5)}=\tau_3$, $T_{\rm max,puo(i)}^{(5)}=T_{\rm max,puo(ii)}^{(5)}=\tau_4$.
   \textbf{Panels (b$_i$)} show the agents' strategies: the green line shows the
   $m_{\mathrm{l}}^{(j_1)(4)}$ in the 4link system. The orange and red lines show the
   $m_{\mathrm{l}}^{(j_1)(5)}$ and $m_{\mathrm{l}}^{(j_2)(5)}$ in the 5link system.
   For comparison the thin, horizontal grey lines show the pure and mixed user optima available
   by externally tuning the agents' decisions. Their values are indicated by the $\sigma_i$
   on the second $y$-axis of panel (b$_5$) with $n_{\rm l,puo}^{(j_1)(4)}=\sigma_3$,
   $\gamma_{\rm muo}^{(4)}=\sigma_3$, $\left(n_{\rm l,puo(i)}^{(j_1)(5)},n_{\rm l,puo(i)}^{(j_2)(5)}
   \right)=\left(\sigma_3,\sigma_1\right)$, $\left(n_{\rm l,puo(ii)}^{(j_1)(5)},
   n_{\rm l,puo(ii)}^{(j_2)(5)}\right)=\left(\sigma_5,\sigma_3\right)$,
   $\left(\gamma_{\rm muo}^{(5)},\delta_{\rm muo}^{(5)}\right)=\left(\sigma_4,\sigma_2\right)$.
   \textbf{Panels (c$_i$)} show the number of switches (i.e. strategy changes) made.
   These numbers are collected in bins with a length of 1000 sweeps. The yellow and red lines
   show how many agents in the 4link system changed strategies and kept their strategies,
   respectively. The blue and green lines show how many agents in the 5link system changed
   and kept their strategies.}
\end{figure*}

In Fig.~\ref{fig:results} we show results for this test state
network with agents choosing their routes intelligently based on the
route choice algorithms described above for different mix-ratios of
information types. The five columns, i.e. panels labelled with
indices 1 to 5, represent different splits of the two information
types as follows: 1: 0\% to 100\%, 2: 25\% to 75\%, 3: 50\% to 50\%,
4: 75\% to 25\%, 5: 100\% to 0\% of users deciding based on
personal-historical and public-predictive information, respectively.
If personal-historical information is used (panels (a-c)$_1$ to
(a-c)$_4$), a certain relaxation process is needed in the algorithm.
The relaxation times are marked by the two vertical lines in the
plots. During the relaxation time, first each agent that relies on
personal historical information tries to use each route at least
once (relaxation step 1 and then 'fills up his memory capacity'
(completing at least $c_{\mathrm{mem}}$ rounds/relaxation step 2. At
the time marked by the black/grey vertical line, the relaxation
procedure in the 4link/5link system is done and the system evolves
according to the route choice algorithm.

Panels (a$_i$) show the time evolution of the average travel time of
the two/three routes. The travel times that are observed in the user
optima with externally tuned decisions are shown for comparison by
the black and grey lines. Their values are given by the $\tau_i$ on
the right y-axes. Panels (b$_i$) denote which routes are chosen by
the agents when deciding based on the route choice algorithm. They
are given by $m_{\rm l}^{(j_1)}$ and $m_{\rm l}^{(j_2)}$, showing
the fractions of agents turning left at junctions $j_1$ and $j_2$.
While they do not represent the strategies of the agents directly,
they can nevertheless be compared to the $n_{\rm l}^{(j_1)}$ and
$n_{\rm l}^{(j_2)}$ or $\gamma$ and $\delta$ of the pure or mixed
user optima attainable by externally tuning all agents' strategies.
One has to keep in mind that the $m_{\rm l}^{(j_1)}$ and $m_{\rm
l}^{(j_2)}$ shown are not directly comparable since individual
intelligently deciding agents do not necessarily keep their
individual strategies. This means that if e.g. the $m_{\rm
l}^{(j_1)}$ and $m_{\rm l}^{(j_2)}$ are identical to the $n_{\rm
l}^{(j_1)}$ and $n_{\rm l}^{(j_2)}$ of an externally tuned pure user
optimum this does not necessarily mean that this user optimum is
realized, since individual agents can still change their strategies
in each round. To get closer to answering the question if a pure or
mixed user optimum is realized, in panels (c$_i$) the number of
switches are shown. They show how many agents switched from one
route to another in the last 1000 time steps (i.e. MC sweeps) and
how many agents kept on the same route. From these observables it
could e.g. be concluded that a pure user optimum is realized if no
switches were made at all. Note that from these quantities it can
not be securely concluded if a mixed user optimum is realized. For
this conclusion the route choice probabilities for each individual
agent would have to be determined.

Panels (a$_1$) to (a$_5$) show that for all splits of information
types in the 4link systems the average travel times of both routes
14 and 23 equalize close to the value that is expected in the
attainable pure and mixed user optima (orange and blue lines) From
panels (b$_1$) to (b$_5$) one can see that at each time
approximately $M/2$ agents are on each of the two routes
($m_{\mathrm{l}}^{(j_1)(4)}\approx 0.5$, green lines). As can be
seen from the green line in panel (b$_i$), if all agents rely on
personal-historical information, oscillations around the user
optimum can be observed. From panels (c$_1$) to (c$_5$) one can see
that most agents stick to one route (red lines) and few agents
switch routes (yellow lines). Summarizing for the 4link we conclude
that for all splits user optimum states are realized in the sense
that average travel times of the two routes equalize.

In the 5link system further interesting effects can be observed.
Panels (b$_1$) to (b$_5$) show that almost no agents choose route 23
($m_{\mathrm{l}}^{(j_1)(5)}\approx 1$, orange lines), and
approximately $M/2$ agents choose routes 14 and 153
($m_{\mathrm{l}}^{(j_2)(5)}\approx 0.5$, red lines). This indicates
that a state close to the second pure user optimum could be realized
(compare orange and red lines to $\left(n_{\rm
l,puo(ii)}^{(j_1)(5)},n_{\rm
l,puo(ii)}^{(j_2)(5)}\right)=\left(\sigma_5,\sigma_3\right)$). It
can be seen that the higher the ratio of public-prescriptive
information gets, the more the value of $m_{\mathrm{l}}^{(j_2)}$
fluctuates around 0.5. This is also confirmed when examining panels
(c$_1$) to (c$_5$): the number of route switches grows with the
amount of users relying on public-predictive information. When all
agents rely on personal-historical information the vast majority of
agents keep their strategies (green line much higher than blue
line). The number of switches grows and for all agents relying on
public-predictive information approximately as many agents change
their routes with each new round as agents stick to their routes.
Regarding the travel times we see, that route 23 which is almost not
used at all has a higher travel time than the two used routes. Their
respective average travel times lie in-between those that are
expected in the accessible pure and mixed user optima. If all agents
rely on personal-historical information they are almost exactly
equal while they are less close to each other if all agents rely on
public-predictive information. Interestingly, their difference does
not grow strictly with the amount of users relying on
public-predictive information as could be assumed by analysing the
strategies and numbers of switches. For all splits the travel times
of the used routes in the 5link system are higher than those of the
routes in the 4link system.

We can thus foremost say that for all splits of information types
Braess states are realized since the 5link travel times are higher
than those in the 4link systems. Furthermore we can summarize that
in the symmetric 4link system with two routes of equal length the
type of information does not seem to have a huge impact: in all
cases situations similar to the attainable optimuma are realized. In
the 5link systems, however, the more public-predictive information
is used the more users keep switching routes from round to round.
This (modern) type of information thus leads to a more unstable
situation (regarding the number of switches). Nevertheless, it does
not seem to lead the system into the attainable mixed user optimum
as the $m_{\mathrm{l}}^{(j_{1/2})(5)}$ are closer to the $n_{\rm
l,puo(ii)}^{(j_{1/2})(5)}$ of the second attainable pure user
optimum.

%%%%%%%%%%%%%%%%%%%%%%%%%%%%%%%%%%%%%%%%%%%%%%%%%%%%%%%%%%%%%%%%%%%%%%%%%%%%%%%%%%

\section{Conclusions}

Our analysis of an exemplary state suggests strongly that Braess'
paradox is likely to still occur in traffic networks in which
drivers choose their routes intelligently based upon information
that is available in modern real-world traffic networks. In our
example, public-predictive information as provided e.g. by
smartphone navigation apps realizes attainable user optima on
average. If all network users base their route choices on their own
past experiences, user optima are also realized in a more stable
way. Also in all mix-ratios the optima are (at least on average)
realized and in all cases Braess states are observed: in the system
with the additional road, users always distribute onto the routes
such that the routes' travel times are higher than those in the
networks without the new road.\\

%%%%%%%%%%%%%%%%%%%%%%%%%%%%%%%%%%%%%%%%%%%%%%%%%%%%%%%%%%%%%%%%%%%%%%%%%%%%%%%%%%%%%%%%%%%
\vspace{0.9cm}

\noindent {\bf Dedication:}

We dedicate this paper to the memory of Dietrich Stauffer. One of
the authors (AS) has known Dietrich for more than 35 years, first as
a student and later as a colleague. He will be remembered not only
for many good advice given through all these years, but also for
lively discussions on a broad range of topics, scientific and beyond
(from politics to football and movies).

\vspace{0.9cm}

\noindent {\bf Acknowledgements:}

Financial support by Deutsche Forschungsgemeinschaft (DFG), Germany
under grant SCHA 636/8-2 and the Bonn-Cologne Graduate School of
Physics and Astronomy (BCGS)is gratefully acknowledged. Monte Carlo
simulations were carried out on the CHEOPS (Cologne High Efficiency
Operating Platform for Science) cluster of the RRZK (University of
Cologne).

%%%%%%%%%%%%%%%%%%%%%%%%%%%%%%%%%%%%%%%%%%%%%%%%%%%%%%%%%%%%%%%%%%%%%%%%%%%%%%%%%%%%%%%%



\bibliographystyle{elsarticle-num}
\bibliography{paper_braess_route_choice}


%\begin{thebibliography}{00}


%\bibitem{1}as

%\end{thebibliography}
\end{document}